\documentclass[prd,aps,floats,twocolumn,superscriptaddress]{revtex4}

\usepackage[dvips]{graphicx}
\usepackage{amssymb}
\usepackage{amsmath}

\usepackage{color}

\begin{document}

\title{Transient solar oscillations driven by primordial black holes}

\author{Michael Kesden} \email{mhk10@nyu.edu}

\affiliation{Center for Cosmology and Particle Physics, Department of Physics,
New York University, New York, New York 10003, USA}

\author{Shravan Hanasoge} \email{hanasoge@princeton.edu}

\affiliation{Department of Geosciences, Princeton University, Princeton,
New Jersey 08544, USA}
\affiliation{Max-Planck-Institut f\"{u}r Sonnensystemforschung,
Max-Planck-Stra$\beta$e 2, 37191 Katlenburg-Lindau, Germany}

\date{September 2011}
                            
\begin{abstract}
Stars are transparent to the passage of primordial black holes (PBHs)
and serve as seismic detectors for such objects.  The gravitational
field of a PBH squeezes a star and causes it to ring acoustically.  We
calculate the seismic signature of a PBH passing through the Sun.  The
background for this signal is the observed spectrum of solar
oscillations excited by supersonic turbulence.  We predict that PBHs
more massive than $10^{21}$ g (comparable in mass to an asteroid) are
detectable by existing solar observatories.  The oscillations excited
by PBHs peak at large scales and high frequencies, making them
potentially detectable in other stars.  The discovery of PBHs would
have profound implications for cosmology and high-energy physics.
\end{abstract}

\maketitle

The existence of black holes is one of the most startling predictions
of general relativity.  Astronomers have discovered two populations of
black holes: stellar-mass ($m_{BH} \sim 10 M_\odot$) black holes that
form in the collapse of massive stars and supermassive black holes
($10^6 M_\odot \lesssim m_{BH} \lesssim 10^{10} M_\odot$) that reside
in galactic centers.  However, general relativity allows black holes
to have any mass.  Black holes much less massive than a solar mass
$M_\odot$ could have formed from density perturbations in the early
Universe.  Such perturbations were created with a wide range of
wavelengths and amplitudes.  Galaxies slowly grew from perturbations
whose amplitudes were initially very small.  Density perturbations
with higher initial amplitudes might have gravitationally collapsed
into primordial black holes (PBHs) \cite{Hawking:1971ei}.  The mass $m_{BH}$ of such PBHs
reflects the mass contained within the particle horizon of the
Universe at the time they were formed.  PBHs as small as the Planck
mass $m_{\rm Pl} = \sqrt{\hslash c/G} \sim 10^{-5}$ g may have formed,
but those with masses less than $m_{\rm evap} \simeq 5 \times 10^{14}$
g would evaporate in less than the age of the Universe
\cite{Hawking:1974rv}.

A density perturbation will collapse into a black hole if its
self-gravity exceeds its pressure support
\cite{Carr:1974nx,Carr:1975qj}.  When this pressure is supplied by
radiation as in the early Universe, PBHs of any mass are equally
likely to form if there is a flat power spectrum of primordial density
perturbations (spectral index $n_s \simeq 1$) as indicated by
observations of the cosmic microwave background \cite{Larson:2010gs}.
PBH production may be greatly enhanced at a particular mass scale if
the pressure were suddenly reduced, such as during the QCD phase
transition \cite{Jedamzik:1996mr}.  The discovery of PBHs of a given
mass would thus provide insight into high-energy physics at the
temperature at which this mass was contained within the particle
horizon.

PBHs are also of great interest to cosmology.  They are collisionless
and nonrelativistic, making them ideal dark-matter candidates.
Observational constraints on the cosmological density $\Omega_{BH}$ of
PBHs depend on their mass $m_{BH}$.  PBHs with masses slightly above
the evaporation limit $m_{\rm evap}$ emit Hawking radiation, including
$\gamma$-rays with a spectrum peaking around 100 MeV
\cite{Page:1976wx}.  Observations of the extragalactic $\gamma$-ray
background by the Energetic Gamma Ray Experiment Telescope (EGRET)
\cite{Sreekumar:1997un} set an upper limit $\Omega_{BH} \leq 5 \times
10^{-10}$ for $m_{BH} = m_{\rm evap}$ \cite{Carr:2009jm}.  Since the
luminosity and temperature of a PBH scale as $m_{BH}^{-2}$ and
$m_{BH}^{-1}$ respectively, it is more difficult to observe Hawking
radiation from larger PBHs.  PBHs of mass $m_{BH} \gtrsim 10^{17}$ g
could constitute the entirety of the dark matter ($\Omega_{BH} =
\Omega_{DM} \simeq 0.23$) without violating observational constraints
on Hawking radiation.

Microlensing surveys constrain the abundance of more massive PBHs.  If
a PBH passes between an observer and a background star, that star will
be gravitationally microlensed, briefly increasing its observed flux.
The duration of each lensing event is proportional to the square root
of the lens mass, implying that a survey with finite cadence will miss
lenses that are too small.  The EROS (Exp\'{e}rience de Recherche
d'Objets Sombres) microlensing survey sets an upper bound of 8\% on
the fraction of the Galactic halo mass in the form of PBHs with masses
in the range $0.6 \times 10^{-7} M_\odot < m_{BH} < 15 M_\odot$
\cite{Alcock:1998fx,Tisserand:2006zx}.

Collisions with Galactic stars might constrain PBHs in this permitted
range $10^{17}~{\rm g} \lesssim m_{BH} \lesssim 10^{26}$ g
\cite{Abramowicz:2008df}.  PBHs passing through a star deposit energy
by dynamical friction.  The fractional loss of kinetic energy $f \sim
10^{-7}(m_{BH}/10^{26}~{\rm g})$ is tiny for PBHs with speeds
comparable to the velocity dispersion of Galactic dark matter.  PBHs
will therefore pass through stars without slowing down or accreting
appreciably.  A $10^{20}$ g PBH was estimated to emit 1 keV x-rays
with a luminosity of a $10^{22}$ erg/s when passing through a
main-sequence star like our Sun \cite{Abramowicz:2008df}, but most of
these x-rays will thermalize and this luminosity is much less than the
Sun's background x-ray luminosity $L_x \gtrsim 10^{26}$ erg/s
\cite{Peres:2000}.

\begin{figure}[t!]
\begin{center}
\includegraphics[width=3.5in]{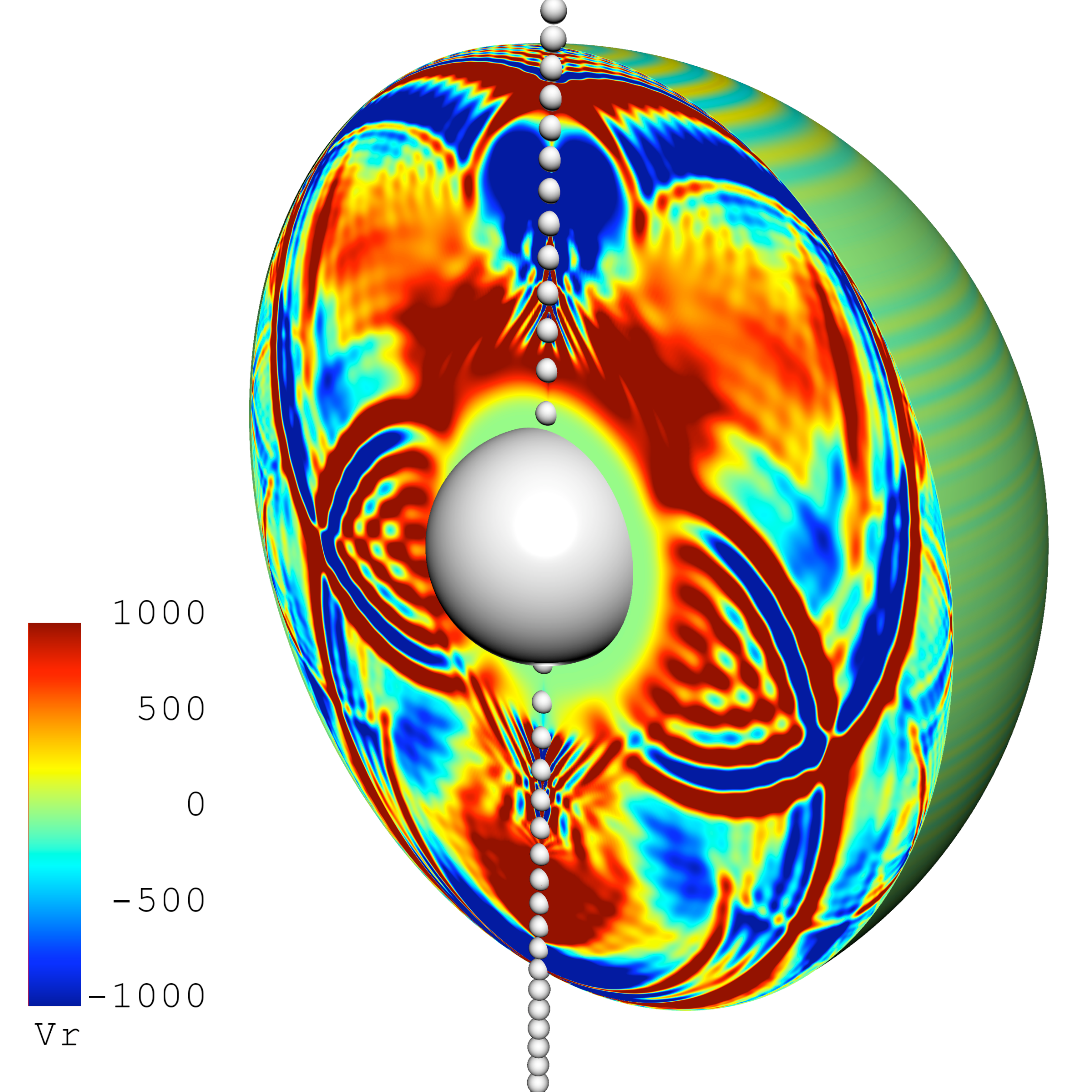}
\end{center}
\caption{Radial velocities $V^{BH}$ induced by a $10^{-10}~M_\odot$ PBH
passing through the Sun along the $z$ axis.  The gray sphere at $R <
0.24 R_\odot$ is excluded from our simulations.  The colors indicate
the radial velocity of the fluid weighted by the square root of the
density.  Wave amplitudes are linearly proportional to the black-hole
mass, so we may rescale our simulations to any desired value of
$m_{BH}$.}
\label{F:RV}
\end{figure}

We instead propose searching for PBHs by the distinctive oscillations
they excite when passing through the Sun.  This signal was considered
previously but not calculated explicitly \cite{Abramowicz:2008df}.  We
simulate the generation and propagation of acoustic waves in the Sun
by solving the linearized Euler equations
\cite{ChristensenDalsgaard:2002ur} in a spherical shell with the
moving PBH acting as the source.  We solve these equations on a
three-dimensional grid consisting of 1024 longitudinal, 512
latitudinal, and 425 radial points extending from $r/R_\odot = 0.24$
to 1.002 \cite{Hanasoge:2006,Hanasoge:2008}.  Our simulations exclude
the core ($r < 0.24~R_\odot$) because of the coordinate singularity at
$r = 0$.  We treat the PBH as a ball with radius $\lambda$ comparable
to our grid spacing.  Although we cannot resolve the PBH's
Schwarzschild radius $R_S = 1.5 \times 10^{-7} (m_{BH}/10^{21}~{\rm
g})$ cm, we verify convergence by repeating simulations with different
values of $\lambda$.  Wavefield velocities are extracted 200 km above
the photosphere $(r = R_\odot)$ to mimic observations.

A snapshot of one of our simulations is shown in Fig.~\ref{F:RV}.  We
can place the PBH's orbit (shown by gray dots) in the $x-z$ plane
without loss of generality because the Sun is nearly spherically
symmetric.  PBH orbits in this plane are fully specified by two
parameters: the energy $E$ and angular momentum $L$ per unit PBH mass.
The PBH orbit in the simulation shown in Fig.~\ref{F:RV} is parabolic
($E = 0$) and radial ($L = 0$).  We consider only unbound orbits ($E
\geq 0$) as gravitationally bound PBHs are extremely unlikely.  The
PBH begins at $z_i \simeq 3 R_\odot$ with an inward velocity of $v_z =
-\sqrt{GM_\odot/z_i}$, falls radially inwards through the Sun's
center, and ends at $z_f \simeq -10 R_\odot$.  The total elapsed time
is 8 h, with the snapshot in Fig.~\ref{F:RV} being taken 7.9 h
into the simulation.  The radial velocity $V^{BH}$ as a function of
time $t$ and angular position $(\theta, \phi)$ on the Sun's surface
constitutes the signal for our proposed PBH search.

\begin{figure}[t!]
\begin{center}
\includegraphics[width=3.5in]{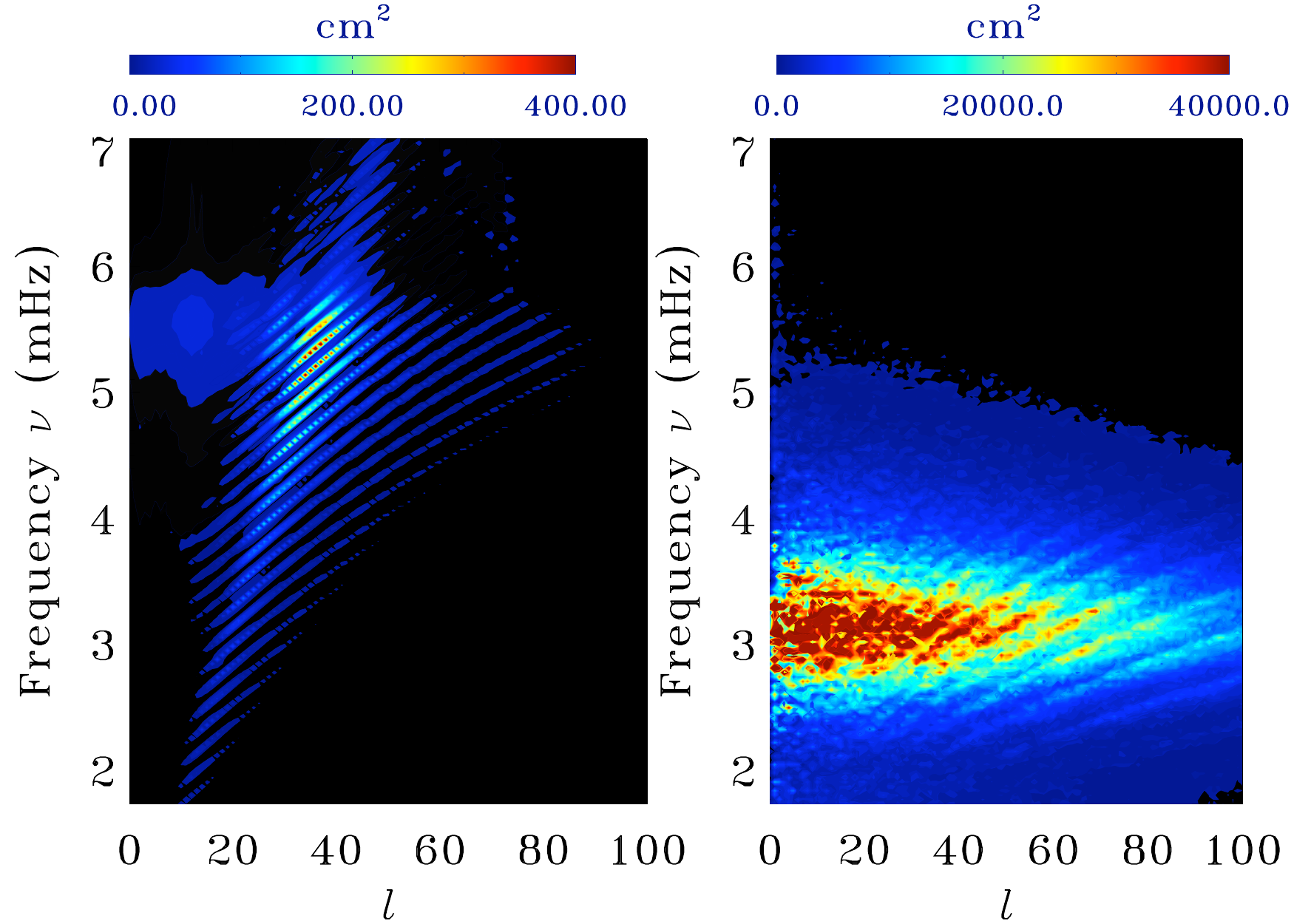}
\end{center}
\caption{Power spectra $P_l(\nu)$ of our simulated PBH signal $V^{BH}$
(left panel) and the turbulently driven noise $V^n$ observed by HMI
(right panel) as functions of frequency $\nu$ and harmonic degree $l$.
Each discrete ridge in both power spectra corresponds to a set of $p$
modes with the same number of radial nodes.  The signal peaks near $l
\simeq 40$, $\nu \simeq 5.5$ mHz, at larger scales and higher
frequencies than the noise background, helping us to distinguish
them.}
\label{F:lnu}
\end{figure}

Photospheric velocities are observed by measuring the Doppler shift of
solar absorption lines.  The Helioseismic and Magnetic Imager (HMI)
\cite{Wachter:2011} onboard the Solar Dynamics Observatory 
currently performs such observations.  The Sun has a discrete spectrum
of global acoustic oscillations known as $p$ modes because pressure
provides the restoring force \cite{ChristensenDalsgaard:2002ur}.  $P$
modes driven by near-surface supersonic turbulence provide the
dominant contribution to observed photospheric velocities, and
constitute the noise background for our PBH search.

\begin{figure}[t!]
\begin{center}
\includegraphics[width=3.5in]{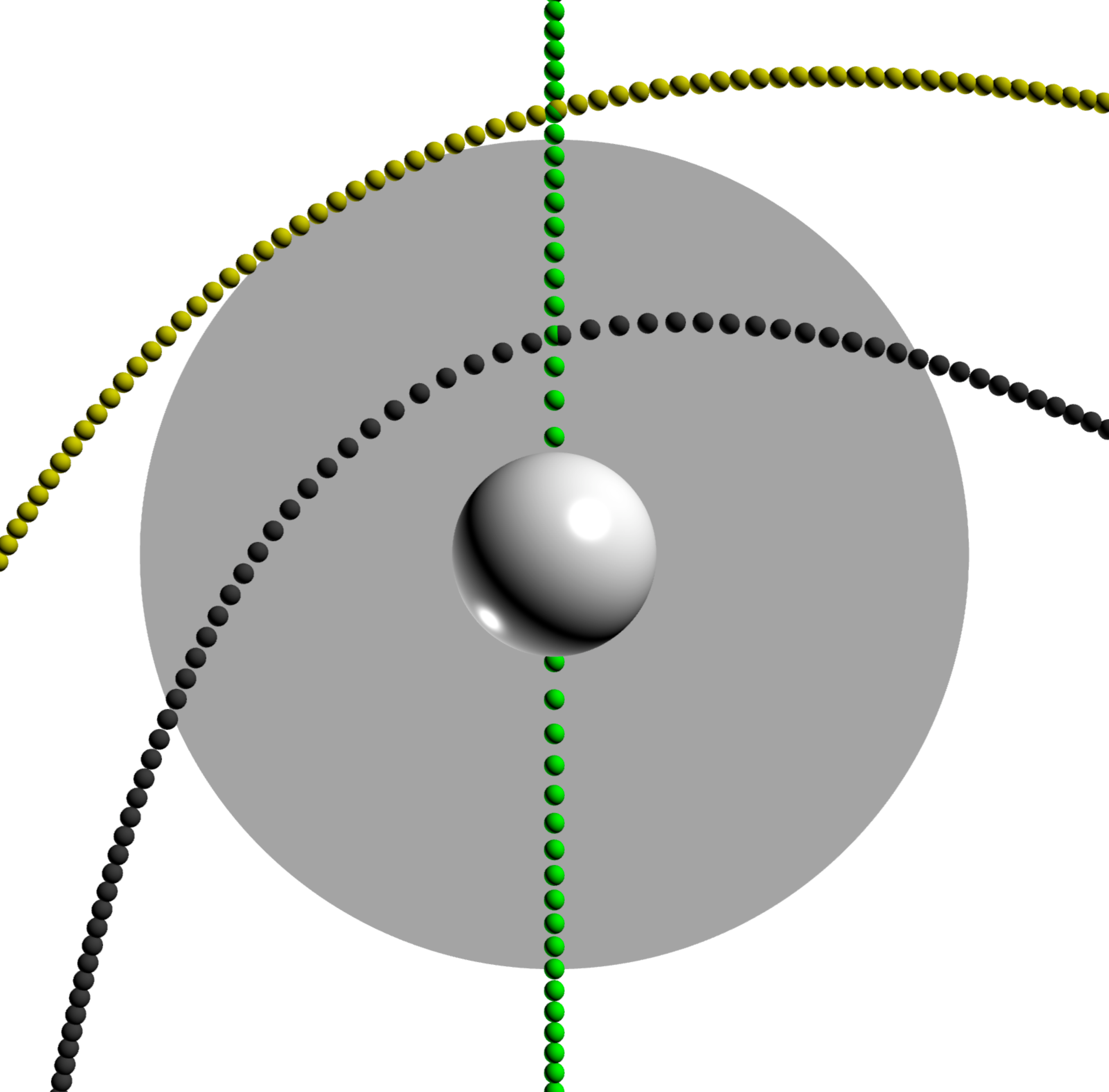}
\end{center}
\caption{Orbits of the PBHs in our simulations.  The green vertical line
along the $z$-axis shows the two radial orbits (RFF and RGC).  The other
two orbits are parabolic ($E = 0$).  The inner black curve shows the
orbit which spends the maximum amount of time within the Sun (MT),
while the outer tan curve shows the orbit which skims the surface
(SS).}
\label{F:orb}
\end{figure}

$P$ modes can be clearly identified after the photospheric velocity
field has been Fourier transformed into a function of frequency $\nu$
and decomposed into spherical harmonics $Y_{lm}(\theta,\phi)$.  We
show the power spectrum
\begin{equation} \label{E:PS}
P_l(\nu) \equiv \frac{1}{2l+1} \sum_{m=-l}^{l} |\tilde{V}_{lm}(\nu)|^2
\end{equation}
of our simulated signal $V^{BH}$ and the observed noise $V^n$ in
Fig.~\ref{F:lnu}.  The power spectrum in the left panel is produced
from the simulation shown in Fig.~\ref{F:RV}, while the noise power
spectrum in the right panel was prepared by using 8 h of publicly
available HMI data \cite{HMI}.  The discrete ridges seen in both
panels correspond to $p$ modes with the same number of radial nodes.
The power spectrum of oscillations excited by the PBH peaks at higher
frequencies and lower harmonic degree $l$ than the noise background
driven by the Sun's natural turbulence.

This dissimilarity between the spatiotemporal dependence of the
signal and noise helps us detect small signals.  We treat each
spherical harmonic of the noise as an independent Gaussian random
variable with variance described by the observed power spectrum
\cite{ChristensenDalsgaard:2002ur}.  When data, such as the observed
velocities $V^{obs}$, are the sum of a signal $V^{BH}$ and a noise
$V^n$, the likelihood that the signal is present in $V^{obs}$ is the
same as the likelihood that $V^{obs} - V^{BH}$ is a Gaussian
realization of the noise in the absence of a signal
\cite{Finn:1992wt}.  This allows us to determine the signal-to-noise
ratio $S/N$ for a given event:
\begin{equation} \label{E:SNR}
\left( \frac{S}{N} \right)^2  = \int_{-\infty}^{\infty} d\nu \sum_l
(2l + 1) \frac{P_{l}^{BH}(\nu)}{P_{l}^{n}(\nu)}~.
\end{equation}
Contributions to the $S/N$ are greatest at harmonic degrees $l$ and
frequencies $\nu$ where the ratio of the numerator in the summation
(left panel of Fig.~\ref{F:lnu}) to the denominator (right panel of
Fig.~\ref{F:lnu}) is maximized.

\begin{figure}[t!]
\begin{center}
\includegraphics[width=3.5in]{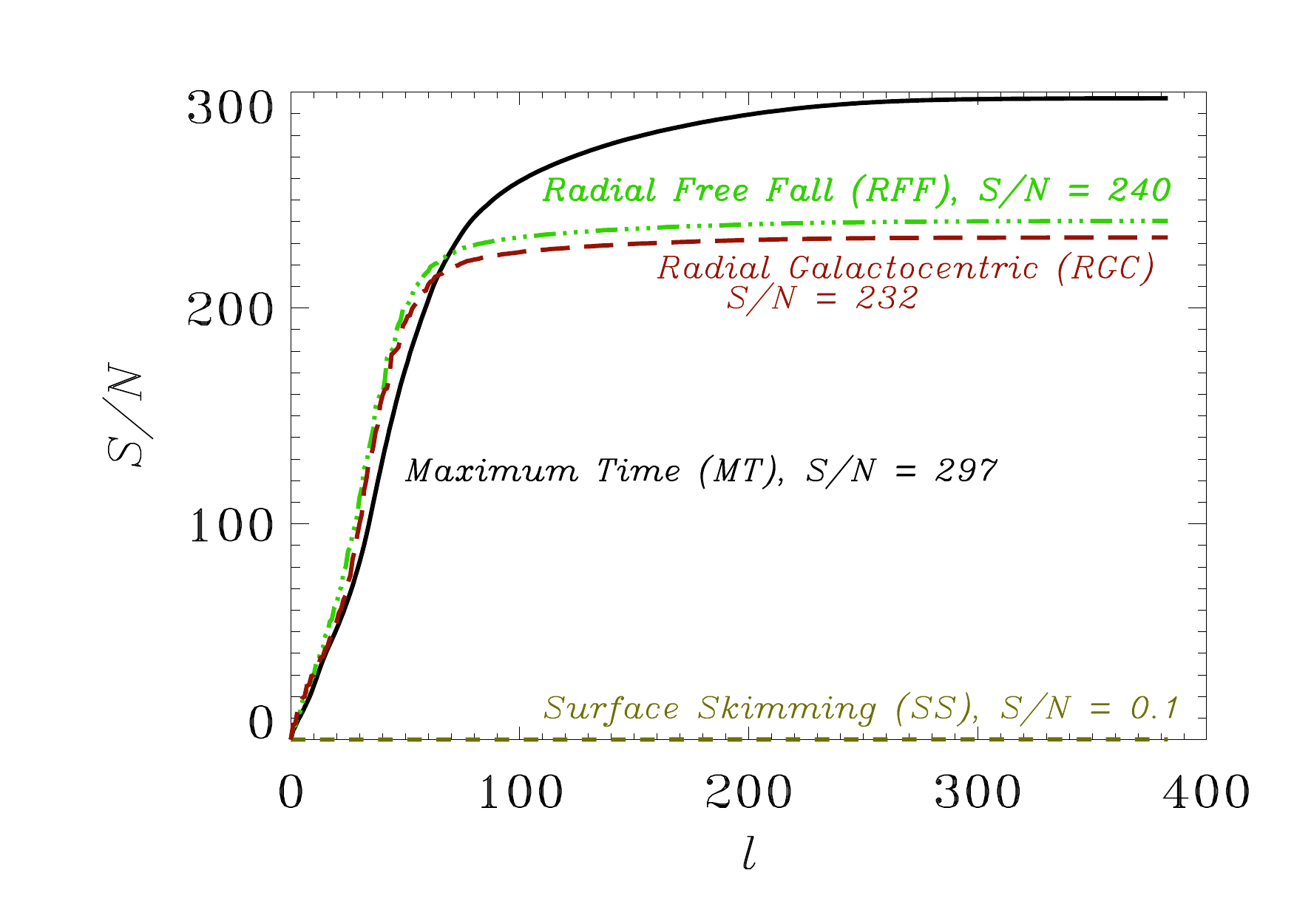}
\end{center}
\caption{Signal-to-noise ratio $S/N$ as a function of maximum
harmonic degree $l$ for $10^{-10} M_\odot$ PBHs on the orbits shown in
Fig.~\ref{F:orb}.  The total $S/N$ for each orbit is listed on each
curve.}
\label{F:SNR}
\end{figure}

A PBH on a radial orbit freely falling from infinity (abbreviated as
RFF hereafter) excites solar oscillations with $S/N = 240$ for $m_{BH}
= 10^{-10} M_\odot$.  This is a conservative estimate for several
reasons.  Our simulations last only for $t_{sim} = 8$ h and thus
neglect contributions to the signal from $t > t_{sim}$.  However, most
of the signal comes from modes with frequencies $\nu$ above the
acoustic cutoff frequency \cite{ChristensenDalsgaard:2002ur}.  These
modes are only partially trapped in the solar interior, escaping into
the corona on the sound-crossing time $\tau \simeq 2\int dR/c_s =
1.96$ h, where $c_s(R)$ is the sound speed as a function of solar
radius.  Simulations with $t_{sim} \gtrsim \tau$ should therefore
capture most of the signal.  Excluding the core ($R < 0.24~R_\odot$)
from our simulations further reduces the signal, by both neglecting
the energy deposited in this region and absorbing waves that reach the
inner boundary.  This absorbing inner boundary condition removes modes
with $\nu \gtrsim 0.2 l$ mHz, as can be seen by the dearth of power in
this region in the left panel of Fig.~\ref{F:lnu}.  The loss of these
modes is significant because the noise has very little power in this
region as seen in the right panel of Fig.~\ref{F:lnu}.  Preliminary
simulations with a core radius of $0.1~R_\odot$ suggest that the true
$S/N$ could be greater by a factor of 2 or more.  These arguments
imply that $S/N = 240$ is indeed a conservative estimate for $m_{BH} =
10^{-10} M_\odot = 2 \times 10^{23}$ g.  Since $S/N \propto m_{BH}$,
the minimum detectable PBH ($S/N \sim 1$) on the RFF orbit will have a
mass $m_{BH} \simeq 10^{21}$ g.

We have performed simulations with PBHs on three additional orbits:
radial infall with a typical galactocentric velocity of 220 km/s at
infinity (RGC), a parabolic orbit that maximizes the time spent at $R
< R_\odot$ (MT), and a parabolic orbit that barely skims the solar
surface (SS).  These orbits are shown in Fig.~\ref{F:orb}.  The $S/N$
for all four simulations as a function of the maximum $l$ included in
the summation in Eq.~(\ref{E:SNR}) is shown in Fig.~\ref{F:SNR}.  Most
of the power comes from large scales ($l \lesssim 100$), since modes
with higher $l$ are evanescent rather than oscillatory deep in the
solar interior where much of the energy is deposited
\cite{ChristensenDalsgaard:2002ur}.  The small difference between the
RFF and RGC simulations, both of which have no angular momentum ($L =
0$), suggests that $S/N$ depends weakly on the orbital energy $E$.
The RFF and MT orbits, both of which are parabolic ($E = 0$),
demonstrate that the $S/N$ depends more strongly on $L$.  Our
preliminary simulations with a core radius $0.1~R_\odot$ suggest an
even stronger dependence on $L$, since the $S/N$ of the RFF simulation
doubles while that of the MT simulation remains nearly unchanged.  The
surface-skimming (SS) orbit has very low $S/N$, implying that the PBH
must penetrate deeply into the solar interior to excite appreciable
oscillations.

One of our primary motivations to search for PBHs is the possibility
that they constitute the cold dark matter required by cosmology.  The
Milky Way resides in a dark-matter halo whose local density is
approximately $\rho_{DM} \simeq 5 \times 10^{-25}$ g/cm$^3$ in the
solar neighborhood.  The Sun orbits the Galactic center with a
velocity $v_\odot \simeq 220$ km/s.  If PBHs with mass $m_{BH}$ and
velocity dispersion $\sigma = v_\odot/\sqrt{2}$ constitute the dark
matter, the differential rate at which PBHs with specific energy $E$
and angular momentum $L$ encounter the Sun is \cite{BT}
\begin{equation} \label{E:Drate}
\frac{\partial^2 N}{\partial E \partial L} =
\frac{2\rho_{DM}L}{m_{BH}v_\odot \sigma} \sqrt{\frac{\pi}{E}}
e^{-(2E + v_{\odot}^2)/2\sigma^2} \sinh \left(
\frac{v_\odot \sqrt{2E}}{\sigma^2} \right)~.
\end{equation}
PBHs with $L < L_{\rm max} = \sqrt{2R_\odot(GM + ER_\odot)}$ have
pericentric distances less than $R_\odot$, implying that the total
rate at which PBHs pass through the Sun is
\begin{eqnarray} \label{E:Rtot}
	N &=& \int_{0}^{\infty} dE \int_{0}^{L_{\rm max}}
	dL~\frac{\partial^2 N}{\partial E \partial L}~, \nonumber \\
	&=& \frac{\pi R_{\odot}^2 \rho_{DM} v_\odot}{m_{BH}} \left[
	\frac{1}{e\sqrt{\pi}} + {\rm erf}(1) \left( \frac{3}{2} +
	\frac{2GM_\odot}{R_\odot v_{\odot}^2} \right) \right]~,
	\nonumber \\
	&=& 10^{-8}~{\rm yr}^{-1} \left(
	\frac{\rho_{DM}}{10^{-25}~{\rm g/cm}^3} \right) \left(
	\frac{m_{BH}}{10^{21}~{\rm g}} \right)^{-1}.
\end{eqnarray}
If all PBHs passing through the Sun were detectable,
Eq.~(\ref{E:Rtot}) would provide the event rate of our proposed
signal.  Since the $S/N$ depends on $m_{BH}$, $E$, and $L$, the true
event rate can be found by weighting the integrand with a Heaviside
step function that vanishes when the $S/N$ is below a chosen
threshold.  Unless there is a considerable enhancement in the local
PBH density beyond that expected for Galactic dark matter, PBHs are
unlikely to be discovered by solar observations alone.

Fortunately, asteroseismologists study oscillations in stars other
than our Sun.  Such oscillations have been observed by the CoRoT
(Convection Rotation and Planetary Transits) \cite{CoRoTAst} and
Kepler satellites \cite{KepAst}.  Resolution limits these instruments
to disk-averaged observations sensitive to only the lowest $l$.
Although this reduces the $S/N$ of each event, the large number of
stars that can be continuously monitored could greatly increase the
total rate of detectable events.  Future missions like the proposed
Stellar Imager \cite{SI} may even resolve stellar disks, allowing
$S/N$ comparable to that for solar events.  Further work is needed to
establish that PBHs can excite detectable oscillations in stars with
structures very different from our Sun.  If particle dark matter is
not detected directly or discovered at the Large Hadron Collider, searches for alternative candidates like PBHs must be
considered.  We believe that asteroseismology might play an important
role in these efforts.

{\bf Acknowledgements.}  All calculations were run on the Pleiades
supercomputer at NASA Ames Research Center.  Many thanks to T. Duvall
Jr. for helping to prepare and interpret HMI observations.  We are
grateful to Tim Sandstrom at NASA Ames for creating impressive
pictures of our simulations including Figs. 1 and 3.  We thank
Glennys Farrar, David Hogg, and Andrew MacFadyen for useful
conversations.  S. H. acknowledges support from NASA Grant
No. NNX11AB63G.

\end{document}